\documentclass[linenumber]{aastex631}
\usepackage{CJK}

\shortauthors{Wei et al.}
\usepackage{graphicx}
\usepackage{amsmath}
\usepackage{txfonts}
\usepackage{threeparttable}
\usepackage{multirow}
\usepackage{threeparttable}
\makeatletter

\newcommand{\Rmnum}[1]{\expandafter\@slowromancap\romannumeral #1@}
\makeatother
\begin{document}
\begin{CJK*}{UTF8}{gbsn}
\title{Intrinsic Pulse Widths of FRB 20121102A and Calculation of Broadening from Propagation and Instrumental Effects }
\correspondingauthor{Yong-Feng Huang}
\email{hyf@nju.edu.cn}
\author{Jia-Peng Wei}
\affiliation{School of Astronomy and Space Science, Nanjing University, Nanjing 210023, China}

\author{Yong-Feng Huang}
\affiliation{School of Astronomy and Space Science, Nanjing University, Nanjing 210023, China}
\affiliation{Key Laboratory of Modern Astronomy and Astrophysics (Nanjing University), Ministry of Education, Nanjing 210023, China}

\author{Lang Cui}
\affiliation{Xinjiang Astronomical Observatory, Chinese Academy of Sciences, 150 Science 1-Street, Urumqi 830011, China}
\affiliation{Key Laboratory of Radio Astronomy, Chinese Academy of Sciences, 150 Science 1-Street, Urumqi 830011, China}
\affiliation{Xinjiang Key Laboratory of Radio Astrophysics, 150 Science 1-Street, Urumqi 830011, China}

\author{Xiang Liu}
\affiliation{Xinjiang Astronomical Observatory, Chinese Academy of Sciences, 150 Science 1-Street, Urumqi 830011, China}
\affiliation{Key Laboratory of Radio Astronomy, Chinese Academy of Sciences, 150 Science 1-Street, Urumqi 830011, China}
\affiliation{Xinjiang Key Laboratory of Radio Astrophysics, 150 Science 1-Street, Urumqi 830011, China}

\author{Jin-Jun Geng}
\affiliation{Purple Mountain Observatory, Chinese Academy of Sciences, Nanjing 210023, China}

\author{Xue-Feng Wu}
\affiliation{Purple Mountain Observatory, Chinese Academy of Sciences, Nanjing 210023, China}

\begin{abstract}
The pulse widths of fast radio bursts are always broadened due to
the scattering of the plasma medium through which the
electromagnetic wave passes. The recorded pulse width will be
further affected by the radio telescopes since the sampling time
and the bandwidth cannot be infinitely small. In this study, we
focus on the pulse widths of at least 3287 bursts detected from
FRB 20121102A as of October 2023. Various effects such as the
scattering broadening, the redshift-induced broadening and the
instrumental broadening are examined. At a redshift of 0.193, the
pulse width is broadened by a factor of approximately 0.16 due to
cosmological time dilation. It is found that the instrumental
broadening only contributes a fraction of $10^{-3}$ -- $10^{-1}$
to the observed pulse width. The scattering broadening is even
smaller, which constitutes a tiny fraction of $10^{-5}$ --
$10^{-2}$ in the observed pulse width. After correcting for these
broadenings, the intrinsic pulse width is derived for each burst.
The maximum and minimum pulse widths at different frequencies are
highlighted. The intrinsic widths of most bursts are in a narrow
range of 1 -- 10 ms, which leads to a quasi-linear correlation
between the fluence and the peak flux. Besides, the mean value and
the dispersion range of intrinsic pulse width are found to scale
with the central frequency as $\nu^{-1.2 \pm 0.2}$ and $\nu^{-1.7
\pm 0.6}$, respectively.
\end{abstract}


\section{Introduction}
\label{sec:intro}

Fast radio bursts (FRBs) are intense radio transients typically lasting for a few milliseconds,
with an isotropic peak luminosity up to $\sim$ $10^{38}$ to $10^{46}$ erg s$^{-1}$. Their brightness
temperature is $\sim$ $10^{35}$ -- $10^{36}$ K, indicating that coherent radiation mechanisms should be
involved (\citealt{Lorimer2007}; \citealt{Cordes&Chartterjee2019}; \citealt{Petroff2019,Petroff2022};
\citealt{Zhang2023}). The dispersion measure (DM) of FRBs, which is an integration of the column density
of free electrons along the line of sight, is generally much larger than the value contributed by
Galactic electrons predicted by the NE2001 model, implying an extra-galactic origin for
them (\citealt{CordesLazio2003}; \citealt{Thornton2013}; \citealt{Yao2017}). There seem to be two
classes of FRBs, i.e., repeaters and non-repeating ones (\citealt{Thornton2013}; \citealt{Spitler2016}; \citealt{Zhang2023}).  Currently, more than 800 FRB sources have
been detected, among which there are only ${\sim}$ 60 FRBs verified to be repeaters and ${\sim}$ 40 are
well-localized (\citealt{Petroff2016,Petroff2022}; \citealt{TheCHIME/FRBCollaboration2021};
\citealt{TheCHIME/FRBCollaboration2022}; \citealt{Hu2023}; \citealt{Xu2023}; \citealt{Zhang2023}).
However, it is still unclear whether those one-off events are really non-repeating sources or are actually also repeaters whose repeated bursts simply were not recorded by us (\citealt{caleb2019}).
Among the repeaters, two sources are found to have long-term periodicity in their activity levels: a possible $\sim$ 160-day periodicity exists
in the activities of FRB 20121102A, and a robust 16.35-day periodicity is found for FRB 20180916B,
with an activity window of ${\sim}$ 5
days (\citealt{Rajwade2020}; \citealt{TheCHIME/FRBCollaboration2020}; \citealt{Cruces2021}).
Statistically, repeating FRBs generally have a relatively larger width and narrower bandwidth (\citealt{andersen2019}; \citealt{Pleunis2021}).
They also have a complex time-frequency down-drifting behavior, generally referred to as
the ``sad-trombone'' effect (\citealt{Hessels2019}; \citealt{TheCHIME/FRBCollaboration2021};
\citealt{Pleunis2021}).

The astrophysical origin of FRBs is still enigmatic. For repeating and one-off FRBs, many
models have been proposed respectively. For example,  possible models of repeating FRBs
include: activities in the magnetosphere of magnetars (\citealt{Beloborodov2017,Beloborodov2020};
\citealt{Margalit2019}), giant pulses from young pulsars (\citealt{Cordes2016}; \citealt{Connor2016}),
or the fractional collapses of the crust of a strange star (\citealt{GENG2021}). On the other
hand, models of non-repeating FRBs include: mergers of double neutron star
systems (\citealt{Totani2013}), or collapses of massive neutron stars to black
holes (\citealt{Falcke&Rezzolla2014}; \citealt{Ravi&Lasky2014}; \citealt{Zhang2018a}).

FRB 20121102A was the first FRB source discovered to repeat, first observed with the 305 m Arecibo telescope (\citealt{Spitler2014}). It has been extensively monitored by many radio telescopes. More than
three thousand FRBs have been observed from this source, which provides a valuable
data set for us to study the nature of FRBs. Some key parameters are available for
the bursts, such as their durations, fluences, and frequency ranges. However, note
that when the radio emissions of FRBs propagate toward us, they will be affected by the
plasma medium along the line of sight, which will lead to various effects such as
the delay of lower frequency waves and the broadening of the pulse width. The plasma
medium is inhomogeneous and includes many components, i.e. the interstellar medium (ISM)
of the host galaxy, the intergalactic medium (IGM), and the ISM of our Milky
Way (\citealt{Rickett1990}; \citealt{Petroff2019}). When the radio waves travel through
the medium, they will be scattered, leading to the delay and broadening of the FRB
pulses (\citealt{Rickett1990}; \citealt{Xu2016}).
Additionally, the radio telescopes may also lead to some broadening of the
observed widths of FRBs (\citealt{Cordes2003}), which is usually referred to as the instrumental effect.
It has been argued that the broadening caused by scattering scales as a function
of frequency as $\tau_{\rm sc} \propto \nu^{-4.4}$
(\citealt{Spitler2016}; \citealt{Michilli2018}; \citealt{Hessels2019}; \citealt{Snelders2023}),
but a detailed analysis on the exact amplitudes of the broadening is still lacking.
In this study, we will investigate the various broadening effects on the at least 3287 bursts
detected from FRB 20121102A by various instruments.

The structure of our article is organized as follows. The theory
of radio wave scattering by plasma and the induced temporal
broadening is introduced in Section \ref{sec:tb}, together with
the introduction of the instrumental broadening. Our numerical
results for 3287 bursts from FRB 20121102A are presented and
analyzed in Section \ref{sec:result}. Finally, our conclusions and
some discussion are included in Section \ref{sec:conclusions}.

\section{Temporal Broadening}
\label{sec:tb}

When a pulse of radio waves travel through a plasma medium, it will be affected
by electron scattering, leading to some chromatic behaviors (dispersion) and the
broadening of the pulse duration. As a result, the recorded pulse width by an
observer can be written as (\citealt{Cordes2003}; \citealt{Lorimer2012}; \citealt{Zhang2023})
\begin{equation}
\label{func1}
W_{\rm obs}  =  [{W_{\rm i}^2}\times(1+z_{0})^2+\tau_{\rm sc}^2+\tau_{\rm tel}^2]^\frac{1}{2} ,
\end{equation}
where $W_{\rm i}$ is the original intrinsic pulse width, $z_{0}$
is the redshift, $\tau_{\rm sc}$ and $\tau_{\rm tel}$ are the
temporal broadenings caused by plasma scattering and instrumental
effects, respectively.

\subsection{Scattering broadening of the pulse width}
\label{2.1}

The total scattering broadening is a combination of the scattering
broadening caused by the ISM in the host galaxy and the Milky Way,
and the IGM (\citealt{Zhang2023}). For simplicity, we assume that
each medium component itself is largely homogeneous and the
intrinsic fluctuations are quasi-statistical (\citealt{Bara1971}).
In this case, the scattering effect can be conveniently modelled
by following the work of \citet{Rickett1990}.

In fact, the scattering broadening caused by IGM is
negligible (\citealt{Katz2016}). Following \citet{Rickett1990}, we
denote the turbulence spectrum index as $\beta$. The typical value
of $\beta$ in case of IGM is 11/3 (i.e. the Kolmogorov turbulence
spectrum). According to \citet{Xu2016}, the outer length scale of
the turbulence medium should be smaller than $\sim 10^{-2}$ pc to
cause a significant scattering broadening with this $\beta$ value.
However, the outer scale of turbulence medium in the IGM is
generally larger than $10^5$ pc (\citealt{Luan2014}). Therefore,
we will neglect the IGM contribution in this study.


For a homogeneous ISM in a galaxy (e.g. the Milky Way or the host
galaxy of a FRB), \citet{Xu2016} argued that the density
fluctuation is likely a short-wave-dominated spectrum ($\beta
<3$). We adopt that the diffractive length is smaller than the
inner length scale of the turbulence $l_0$ and the scattering
angle is much less than 1. Here we replace their $\delta n_{\rm
e}$ with $\sqrt{f}\delta n_{\rm e}$, where $f$ is the volume
filling factor and $\delta n_{\rm e}$ is the root-mean-square of
the electron density fluctuation (\citealt{Xu2016}).  Therefore,
we can derive the broadening timescale induced by the scattering
of the galactic ISM  as
\begin{equation}\label{func2}
\tau_{\rm sc}  \propto
\frac{\lambda_0^4 \mathrm{DM}^2 f}{(1+z_0)^3}(\frac{\delta n_{\rm e}}{n_{\rm e}})^2 l_0^{-1},
\end{equation}
where DM is the dispersion measure of the corresponding
ISM (it can be either the ISM of the Milky Way or the ISM
of the host galaxy), $\lambda_0$ is the wavelength of the radio
emission in the observer's frame, and $n_{\rm e}$ is
correspondingly the electron number density.  The density contrast
in the host galaxy of FRB 20121102A is unknown. Here we take
$\delta n_{\rm e} / n_{\rm e} \sim 1$ in our calculations, which
is typical in the Milky Way (\citealt{Xu2016}). Equation
(\ref{func2}) can then be further expressed as
\begin{equation}
\label{func3}
\tau_{\rm sc}  \approx  \frac{0.2}{(1+z_{0})^3}(\frac{\mathrm{DM}}{100\, {\rm pc\, cm^{-3}}})^2(\frac{\lambda_{0}}{{\rm 1m}})^4
(\frac{f}{10^{-6}})(\frac{\delta n_{\rm e}}{n_{\rm e}})^2(\frac{l_{0}}{10^{-10}{\rm pc}})^{-1} {\rm ms},
\end{equation}
by substituting the typical values of the parameters into it (\citealt{Xu2016}).
In our calculations, we take $f \sim 10^{-6}$ and $l_0 \sim 10^{-10}$ pc (\citealt{Xu2016}).
We then can use this equation to conveniently
calculate the contribution of scattering broadening from the ISM
of the host galaxy and the Milky Way. For example, by
substituting the redshift ($z_0$) and dispersion measure of the
host galaxy ($\mathrm{DM}_{\rm host}$), and the observing
frequency into Equation (\ref{func3}), we can obtain the
scattering timescale due to the ISM of the host galaxy. We can
also calculate the scattering timescale caused by the ISM of the
Milky Way similarly.

However, $\mathrm{DM}_{\rm host}$ cannot be directly
measured through observations. Here we estimate it as follows.
First, the total dispersion measure on the line of sight is a sum
of several components (\citealt{Thornton2013}; \citealt{Deng2014};
\citealt{Prochaska2019}), i.e.
\begin{equation}\label{func4}
\mathrm{DM}_{\rm total} = \mathrm{DM}_{\rm MW} + \mathrm{DM}_{\rm halo} + \mathrm{DM}_{\rm IGM} + \frac{\mathrm{DM}_{\rm host}}{1+z_{0}},
\end{equation}
where $\mathrm{DM}_{\rm MW} ,\, \mathrm{DM}_{\rm
halo}\, \mathrm{and}\, \mathrm{DM}_{\rm IGM}$  are the
contributions from the Milky Way, the Galactic halo and the IGM,
respectively. Note that $\mathrm{DM}_{\rm total}$ can
be derived from FRB observations. Second, for our Galaxy, we have
$\mathrm{DM}_{\rm MW} = 188 \, {\rm pc\, cm^{-3}}$
and $\mathrm{DM}_{\rm halo} = 30 \, {\rm pc\,
cm^{-3}}$ in the direction of FRB 20121102A by referring to the NE2001 Galactic ISM model (\citealt{CordesLazio2003};
\citealt{Zhang2023}). At the same time, the IGM contribution at
$z_{0} < 3$ can be calculated by using a linear formula of
(\citealt{Zhang2018a}; \citealt{Pol2019}; \citealt{cordes2021})
\begin{equation}\label{func5}
 <\mathrm{DM}_{\rm IGM}>  =  (855\,{\rm pc\,cm^{-3}})z_{0}\frac{H_{0}}{67.74\,{\rm km\,s^{-1}\,kpc^{-1}}}
\frac{\Omega_{\rm b}}{0.0486}\frac{f_{\rm IGM}}{0.83}\frac{\chi}{7/8},
\end{equation}
where $H_{0}$ is the Hubble constant, $\Omega_b$ is the energy
density fraction of baryons, $f_{\rm IGM}$ is the mass fraction of
baryons in the IGM, and $\chi$ is the fraction of ionized
electrons. So, from Equation (\ref{func4}), we can get an
estimation of $\mathrm{DM}_{\rm host}$ by subtracting other
components from $\mathrm{DM}_{\rm total}$. In this way,
$\mathrm{DM}_{\rm host}$ of FRB 20121102A is estimated to be in a
range of 200 -- 220 $\mathrm{pc\ cm^{-3}}$. Consequently, the
scattering broadening caused by the ISM of the host galaxy ranges
from $10^{-3}$ ms to $10^{-5}$ ms, which is comparable to the ISM
contribution of the Milky Way.

\subsection{Instrumental broadening}
\label{2.2}

The radio telescope can also have a complex effect on the observed pulse width.
It depends on the DM of the source, the bandwidth of each frequency channel,
and the sampling time. To be more specific, the instrumental broadening term
of $\tau_{\rm tel}$ in Equation (\ref{func1}) can be written as (\citealt{Cordes2003}; \citealt{Petroff2019})
\begin{equation} \label{func6}
\tau_{\rm tel}  =  [\Delta\tau_{\rm DM}^2+\Delta\tau_{\rm \delta DM}^2+\Delta\tau_{\Delta\nu}^2+\tau_{\rm samp}^2]^\frac{1}{2} .
\end{equation}
Here, the first term, $\Delta\tau_ {\rm DM}$, is the frequency-dependent intra-channel
dispersive smearing, which is further expressed as (\citealt{Cordes2003})
\begin{equation}\label{func7}
\Delta\tau_{\rm DM}  =  8.3\mathrm{DM}\Delta\nu_{\rm MHz}\nu_{\rm GHz}^{-3}\,\,\mu {\rm s},
\end{equation}
where $\Delta\nu_{\rm MHz}$ is the bandwidth of the channel (in units of MHz)
and $\nu_{\rm GHz}$ is its central frequency (in units of GHz).

The second term in Equation (\ref{func6}) originates from the deviations
(i.e. $\delta  \mathrm{DM}$) between the true DM and the fiducial DM, which is chosen
to de-disperse the channels coherently (\citealt{Hessels2019}). It can be
expressed as (\citealt{Cordes2003})
\begin{equation}\label{func8}
\Delta \tau_{\rm \delta DM}  = {\Delta\tau_{\rm DM}}\frac{\delta \mathrm{DM}}{\mathrm{DM}}.
\end{equation}
The third term in Equation (\ref{func6}) is the so-called bandwidth
smearing (\citealt{Bridle1999}; \citealt{Rioja2018}), which can be approximately
calculated as
\begin{equation}\label{func9}
\Delta \tau_{\Delta \nu} \sim (\Delta \nu_{\rm MHz})^{-1} \,\,\mu {\rm s}.
\end{equation}
Finally, the fourth term of $\tau_{\rm samp}$ in Equation (\ref{func6}) is
simply the sampling time.

\subsection{Redshift-induced broadening}

The cosmological time dilation can also cause the
broadening of pulse widths, namely, the redshift-induced
broadening. It can be expressed as
\begin{equation}\label{func10}
    \tau_{\rm red} = W_{\rm i}\times z_{0}.
\end{equation}
Note that the intrinsic pulse width of $W_{\rm i}$ in Equation
(\ref{func10}) cannot be directly measured from observations, so
this equation cannot be straightforwardly applied. Instead, we
need to solve Equation (\ref{func1}) to derive $W_{\rm i}$ first,
which can be done by substituting the scattering broadening and
instrumental broadening terms as described in Subsections
\ref{2.1} and \ref{2.2} into Equation (\ref{func1}). Consequently,
utilizing the derived intrinsic pulse widths of FRBs, we can
calculate the redshift-induced broadening by using Equation
(\ref{func10}). In the case of FRB 20121102A which resides at $z =
0.193$, since the scattering broadening and instrumental
broadening are generally insignificant (see the section below), we
find that the pulse width is broadened by a factor of
approximately 0.16 due to cosmological time dilation. 

\section{Numerical Results for FRB 20121102A}
\label{sec:result}

As the first repeating FRB source,
FRB 20121102A has been extensively monitored by many radio telescopes. The source
is associated with a low-metallicity dwarf galaxy at a redshift of ${\sim}$ 0.193,
located in the vicinity of a star-forming region (\citealt{Cordes2006};
\citealt{Spitler2014,Spitler2016}; \citealt{Lazarus2015}; \citealt{Chatterjee2017};
\citealt{Tendulkar2017}; \citealt{Marcote2017}). The DM of FRB 20121102A was
measured as 557.4 ${\pm}$ 2.0 pc cm$^{\rm-3}$ on MJD 56233, and has increased to
565.8 ${\pm}$ 0.9 pc cm$^{\rm-3}$ between MJD 58724 and MJD 58776, showing an
increasing trend of ${\sim}$ 1 pc cm$^{\rm-3}$ year$^{\rm-1}$
(\citealt{Spitler2014,Spitler2016}; \citealt{Li2021}). A linear polarization of
nearly 100 percent was detected, with the polarization angle being almost
constant (\citealt{Michilli2018}; \citealt{Plavin2022}; \citealt{Hewitt2022}).
In the meantime, the rotation measure (RM) of FRB 20121102A has changed drastically
from 1.46 ${\times}$ 10$^{\rm5}$ rad m$^{\rm-2}$ to 7 ${\times}$ 10$^{\rm4}$
rad m$^{\rm-2}$ over three years, which could point to the existence of an
extreme magneto-ionic environment related to an accreting compact star,
or a magnetized wind nebula around a magnetar (\citealt{Spitler2016};
\citealt{Michilli2018}). In addition, other environment models have
been proposed, such as the binary interactions involving a neutron
star (\citealt{Wang2022}). As of October 2023, at least 3287 bursts have been detected from FRB 20121102A by various telescopes (\citealt{Li2021}; \citealt{Aggarwal2021}; \citealt{Hewitt2022}; \citealt{Jahns2023}). We have
collected the main observational parameters of these events from previous literature.
An overview of our data set is presented in Table \ref{table1}. In our study, we have directly taken the
observed parameters from the corresponding references.

\begin{figure*}
        \centering
        \includegraphics[width=0.55\textwidth]{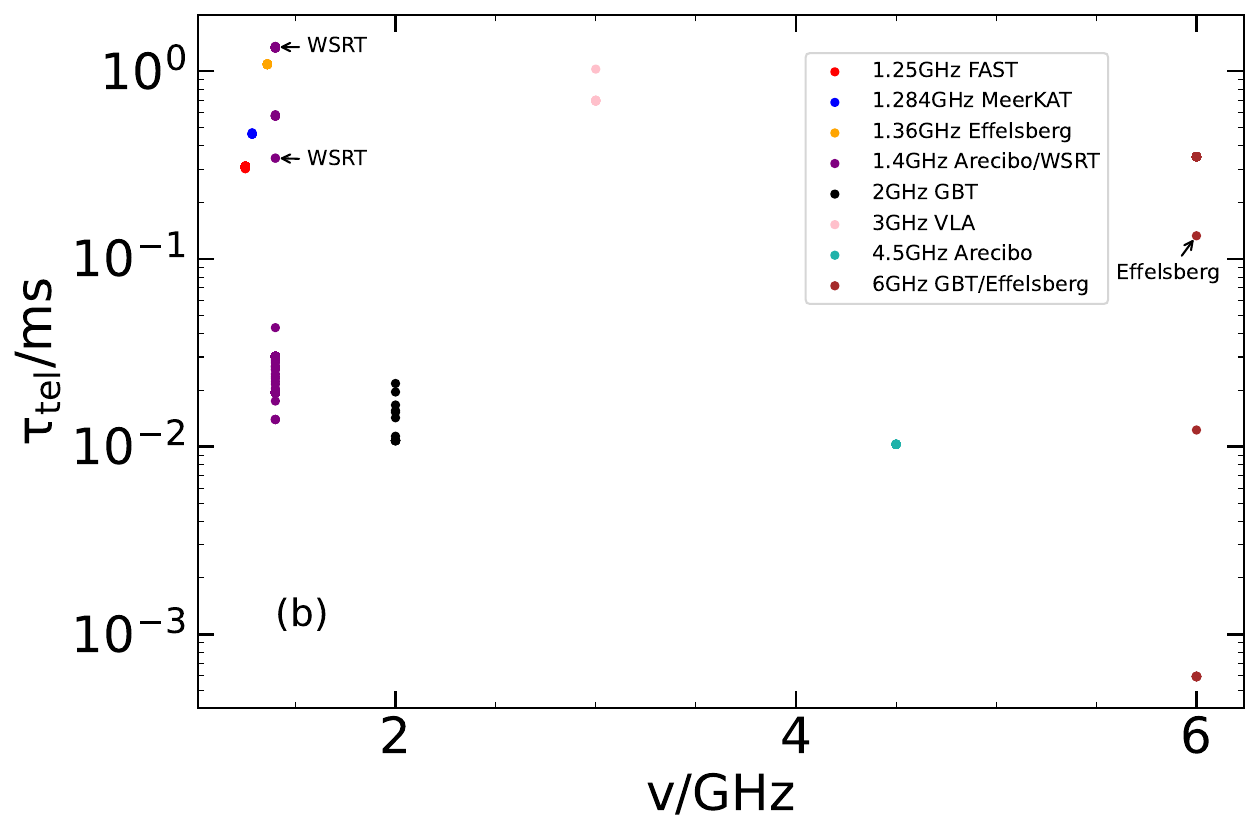}
    \caption{
     The instrumental broadening plotted versus the central frequency.
             This broadening effect is calculated by using Equation (\ref{func6}).
            }
    \label{fig1}
\end{figure*}

For the 3287 bursts from FRB 20121102A collected in our sample, we
have comprehensively analyzed the broadening of their pulse widths
caused by different factors by using the equations introduced in
Section \ref{sec:tb}. For example, the scattering
broadening can be calculated with Equation (\ref{func3}). It is
found that the total scattering timescale $\tau_{\rm sc}$ caused
by both the Milky Way and the host galaxy ranges from $10^{-5}$ to
$10^{-3}$ ms. The effect of instrumental broadening at different
observing frequencies is illustrated in Figure \ref{fig1}. We see
that $\tau_{\rm tel}$ is generally between $10^{-2}$ ms to 1 ms,
which is about three magnitudes larger than the scattering
broadening. This effect is significant for short FRBs,
especially those lasting for less than 1 ms.

\begin{figure}
        \centering
        \includegraphics[width=0.55\textwidth]{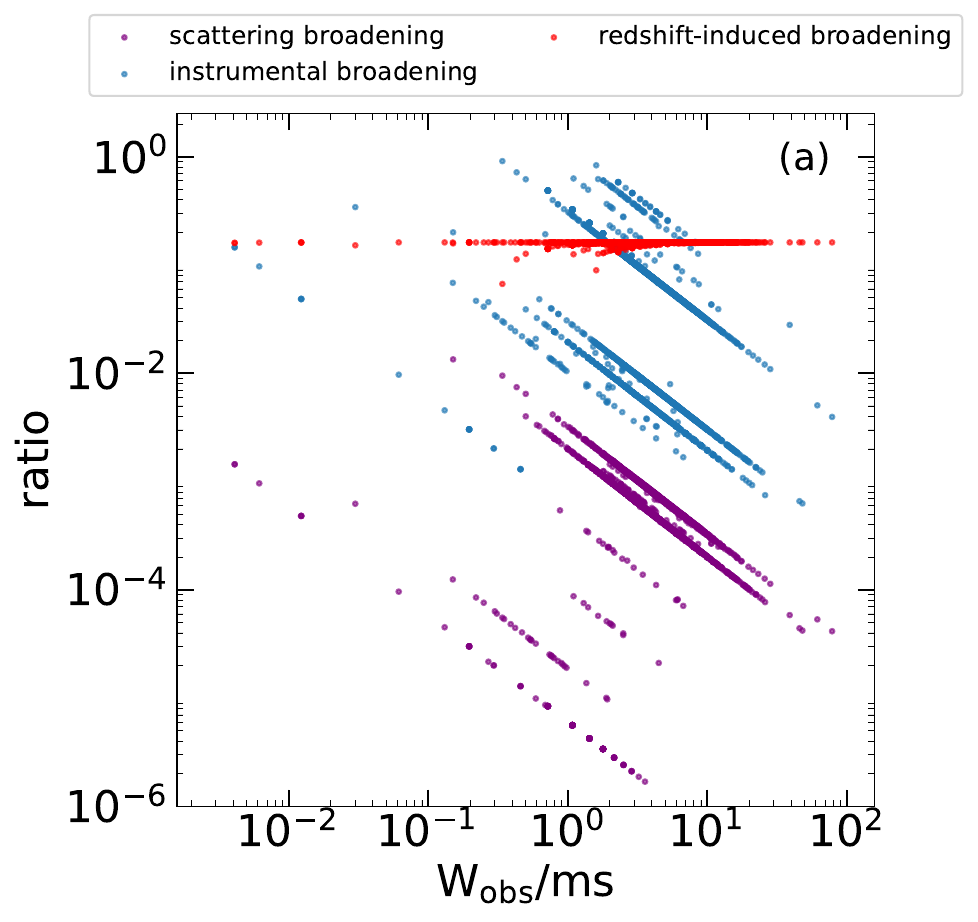}
        \includegraphics[width=0.49\textwidth]{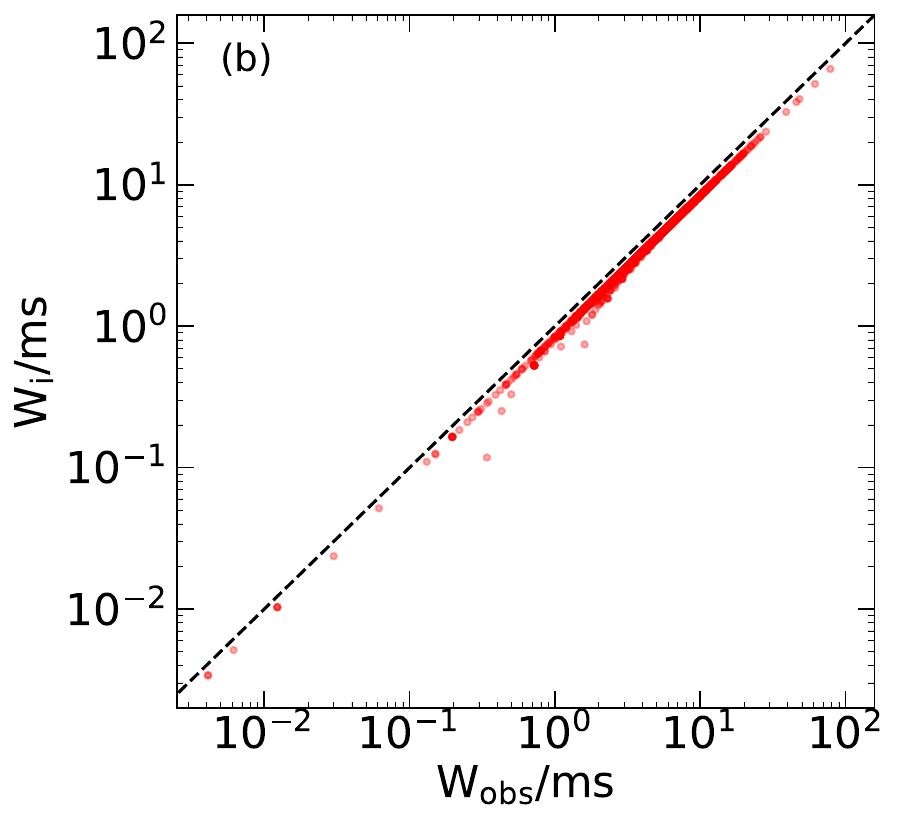}
        \includegraphics[width=0.49\textwidth]{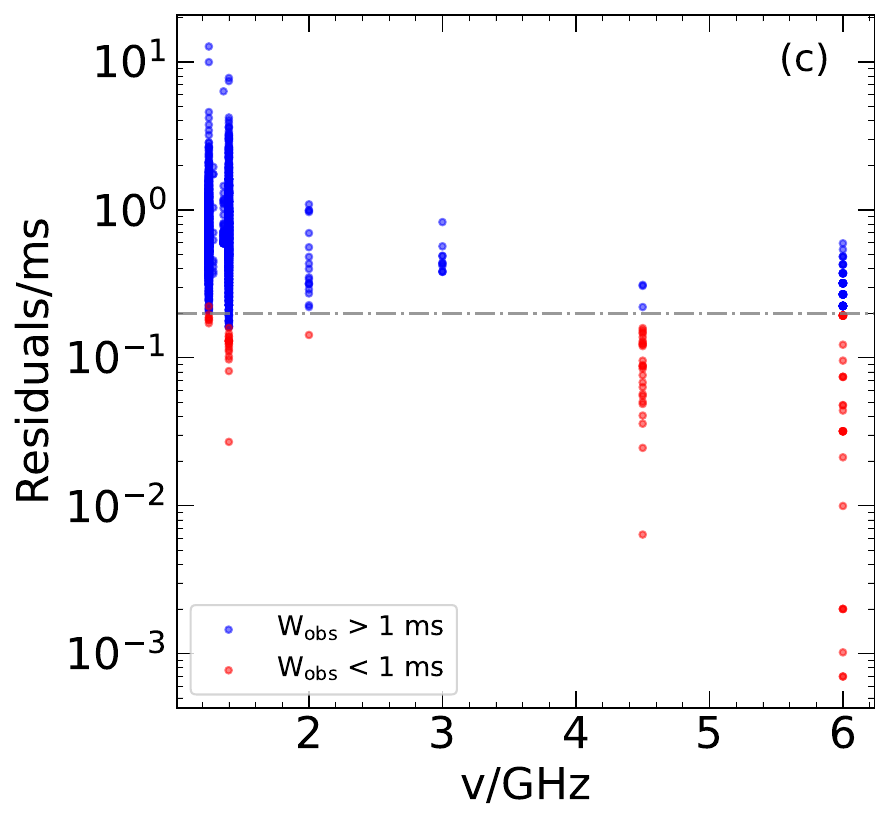}
        \caption{
         Various broadening effects on the bursts of FRB 20121102A.
        The first panel shows the ratio of pulse width broadening due to
        the redshift (${\tau_{\rm red}}/{W_{\rm obs}}$),
        scattering (${\tau_{\rm sc}}/{W_{\rm obs}}$) and
        instrumental (${\tau_{\rm tel}}/{W_{\rm obs}}$) effects (marked
        by different colors). Note that there are three points for each
        burst (one for each category of broadening).
        The second panel plots the intrinsic pulse width versus the
        observed width. The dashed line corresponds to $W_{\rm obs} = W_{\rm i}$.
        The third panel plots the residuals between observed and
        intrinsic pulse widths versus the central frequency. The gray
        dash-dotted line corresponds to a residual of 0.2 ms. The blue points
        indicate the bursts with $W_{\rm obs}$ larger than 1 ms,
        while the red points correspond to bursts with $W_{\rm obs}$ less than 1 ms.}
    \label{fig2}
\end{figure}

In Figure \ref{fig2}a, the fractions of pulse width
broadening due to the redshift ($\tau_{\rm red}/W_{\rm obs}$),
scattering ($\tau_{\rm sc}/W_{\rm obs}$) and instrumental
($\tau_{\rm tel}/W_{\rm obs}$) effects are shown for each pulse.
The three effects are illustrated by different colors.
Since FRB 20121102A is at a redshift of $z \sim 0.193$, we
see that the redshift-induced broadening ratio is generally $\sim
0.16$ (the red dots). It leads to a broadening of $\sim 0.1 $ --
1 ms for most FRBs. The ratio of instrumental broadening generally
ranges between $10^{-3}$ -- 0.1. However, for some bursts, the
ratio can be larger than 60\%. They are mainly short events whose
duration is less than 1 ms. In these cases, the sampling time
could seriously affect the measured pulse width. Furthermore, we
see that the ratio of scattering broadening generally ranges
between $10^{-5}$ -- $10^{-2}$.  It is much smaller than the
instrumental broadening as a whole. After solving the effects of
these factors, we can then easily correct them and derive the
intrinsic pulse width, i.e. $W_{\rm i}$. Figure \ref{fig2}b plots
$W_{\rm i}$ versus the observed width. Here, the dashed line
corresponds to $W_{\rm i} = W_{\rm obs}$. We see that most of the
data points are very close to the dashed line, which means that
the effects of all three broadening factors (i.e. redshift,
scattering and instrumental effects) are unimportant for the
majority of FRBs. The broadening is noticeable only for a very
small portion of FRBs. They are typically short events that are
more seriously subjected to the influence of the sampling time.
Figure \ref{fig2}c plots the residuals between the
observed and intrinsic pulse widths versus the central
frequencies. Here the residual equals to $W_{\rm
obs}[1-(1-\tau_{\rm sc}^2/W_{\rm obs}^2-\tau_{\rm tel}^2/W_{\rm
obs}^2)^{\frac{1}{2}}/(1+z_0)]$. The gray dash-dotted line
corresponds to a residual of 0.2 ms. We see that for the bursts
with the width longer than 1 ms, the residual is generally larger
than 0.2 ms, while for those shorter than 1 ms, the residual is
generally less than 0.2 ms. At a particular frequency, the
variation of the residual is mainly due to the difference of the
observed width, i.e. $W_{\rm obs}$. 

Figures \ref{fig1} -- \ref{fig2} present an overall
analysis of all the pulses in our sample. Next, we will divide
the 3287 bursts into subsamples according to the radio telescopes
that detected them and analyze these subsamples separately.
In our study, the $\texttt{curve-fit}$ and
$\texttt{scipy.stats}$
codes$\footnote{https://docs.scipy.org/doc/scipy/reference/index.html}$,
which are based on the least square fitting method, are adopted to
get a best-fit curve for the data points. Monte Carlo simulations
are performed to obtain the corresponding confidence level of the
fitting curve.

\begin{figure}
        \centering
    \includegraphics[width=0.49\textwidth]{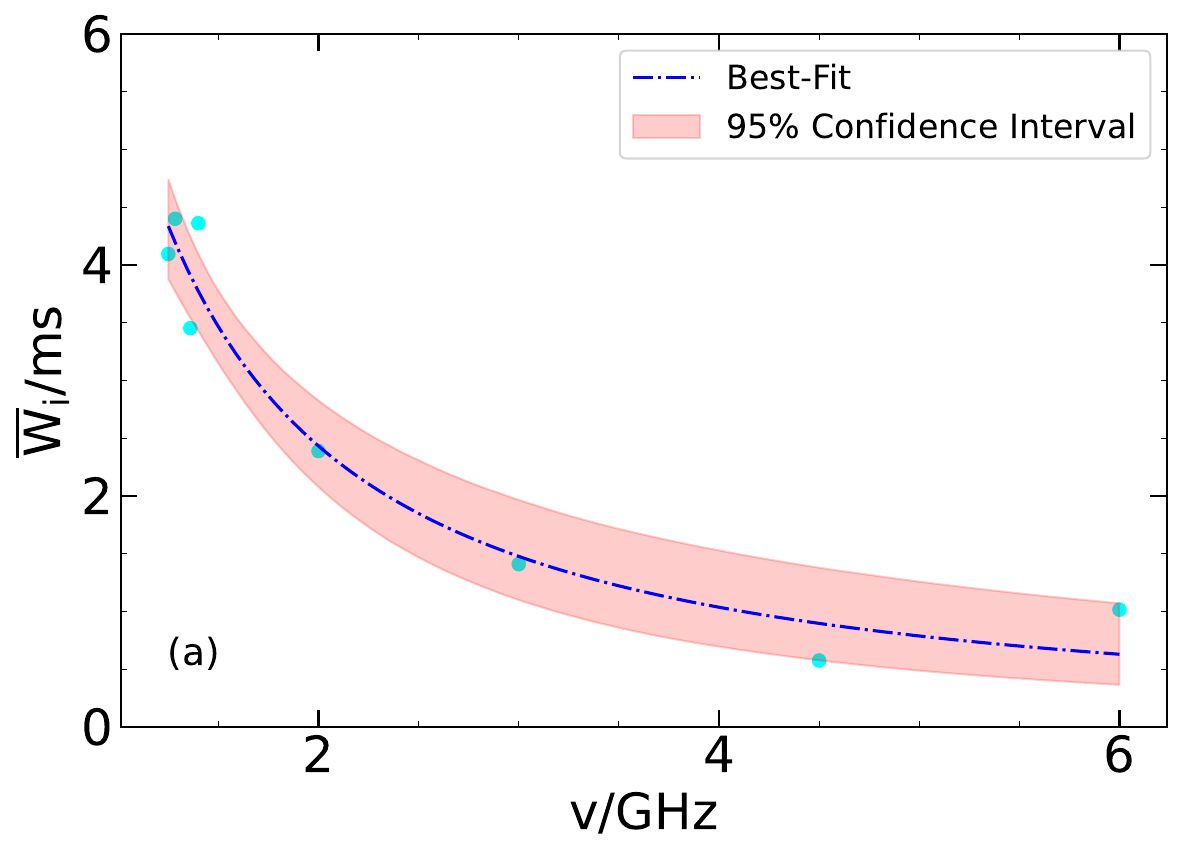}
        \includegraphics[width=0.49\textwidth]{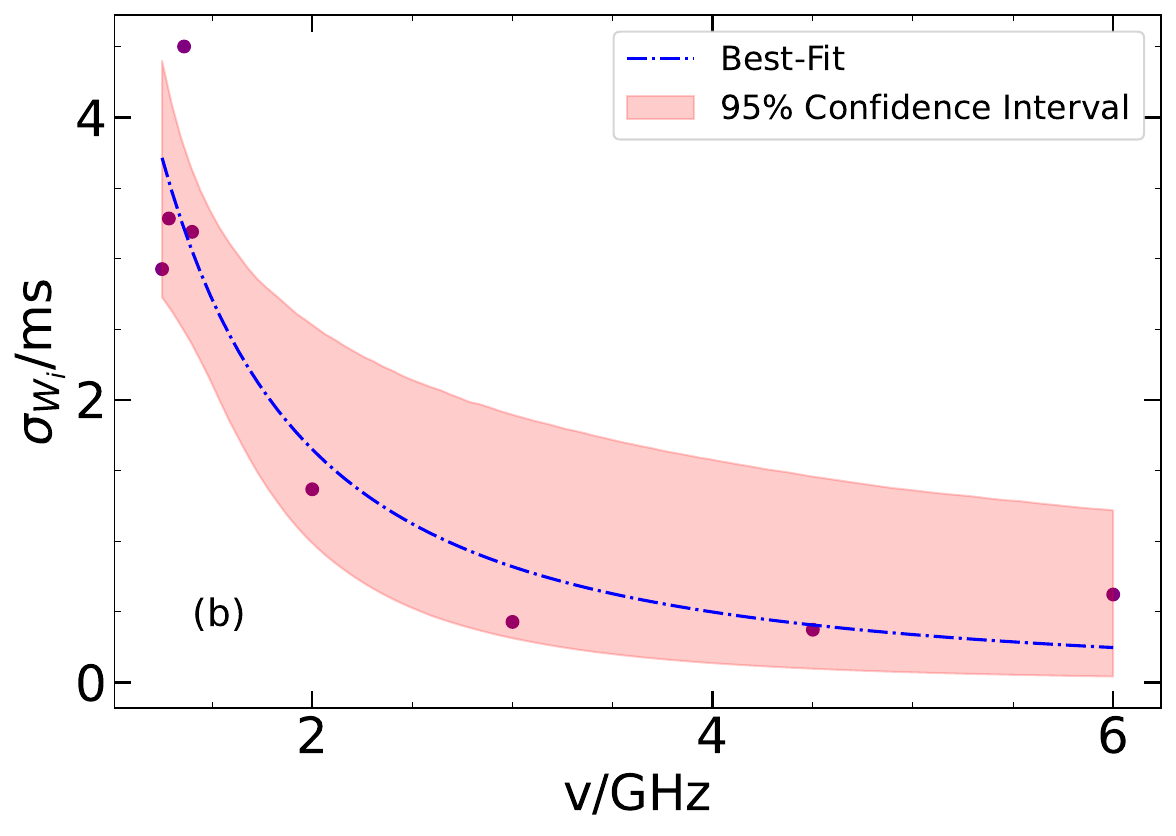}
        \includegraphics[width=0.49\textwidth]{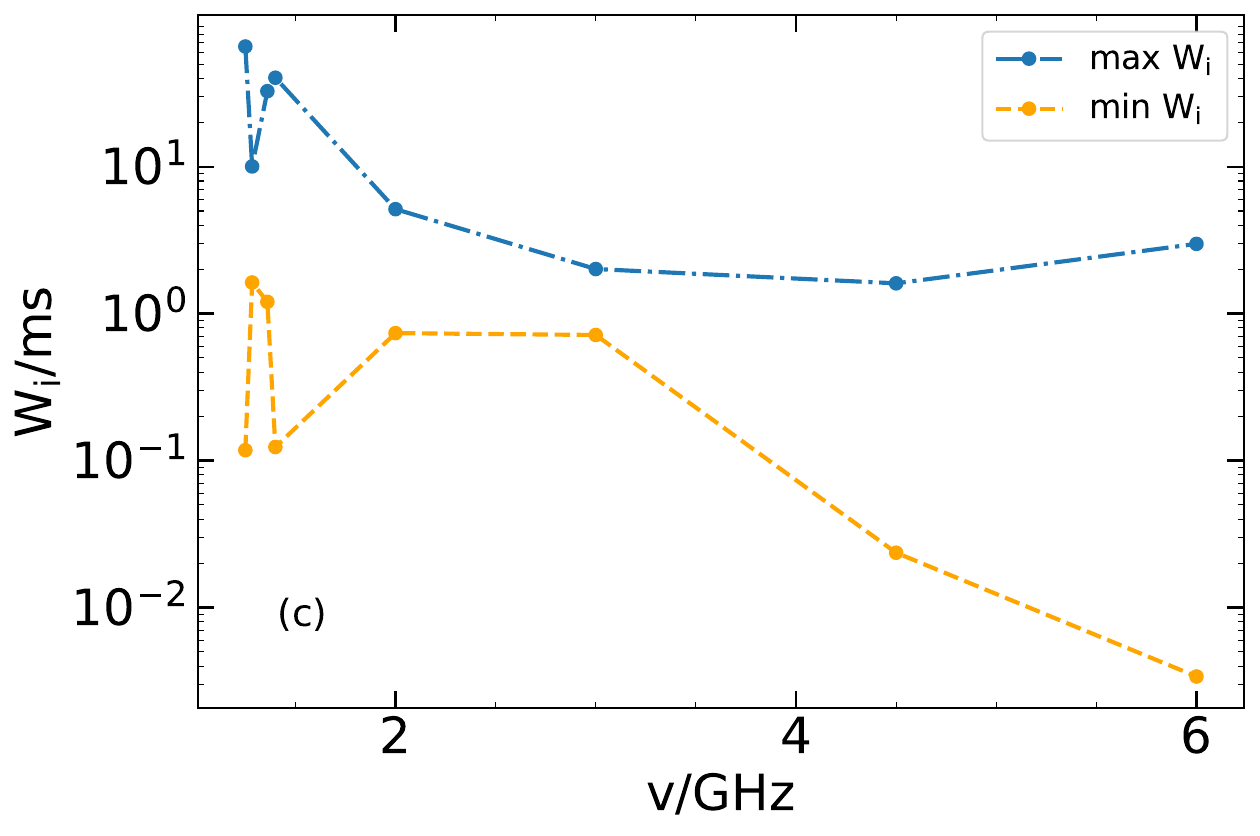}
        \caption{Characteristics of the intrinsic pulse widths based on
    subsample analysis. Here the FRBs are divided into subsamples according
    to the telescopes that detected them. Panel (a): the mean intrinsic
    pulse width of each subsample ($\overline{W}_{\rm i}$) plotted versus
    the central frequency; Panel (b): the dispersion range of the intrinsic
    pulse width ($\sigma_{\rm W_i}$) plotted versus frequency for each
    subsample; Panel (c): the maximum and minimum values of the intrinsic
    pulse width in each subsample versus frequency.}
    \label{fig3}
\end{figure}

Our results based on subsample analysis of the intrinsic pulse
width are illustrated in Figure \ref{fig3}. Figure \ref{fig3}a
plots the mean intrinsic pulse width ($\overline{W}_{\rm i}$)
versus the central frequency of the FRBs. We see that
$\overline{W}_{\rm i}$ obviously has a decreasing tendency when
the frequency increases, which has been observed by other
groups as well (\citealt{Gajjar2018}; \citealt{Chamma2023}). It
declines from  $\sim 4.4$ ms at 1.3 GHz to $\sim 1$ ms at 6 GHz
and generally follows a function of $\overline{W}_{\rm i} \propto
\nu^{-1.2 \pm 0.2}$. In fact, the best-fit curve of the data
points corresponds to

\begin{equation}\label{func11}
    \frac{ \overline{W}_{\rm i}}{1 {\rm ms} } = (5.7\pm 0.5) \left( \frac{\nu}{1 {\rm Ghz}} \right)^{-1.2 \pm 0.2}.
\end{equation}

Figure \ref{fig3}b plots the dispersion range of the intrinsic pulse width ($\sigma_{\rm W_i}$)
versus the central frequency. Interestingly, $\sigma_{\rm W_i}$ also
has a decreasing tendency when the frequency increases. It declines from  $\sim 4.5$ ms
at 1.3 GHz to $\sim 1$ ms at 6 GHz, and scales as
$\sigma_{\rm W_i} \propto \nu^{-1.7 \pm 0.6}$. The best-fit curve of the data
points corresponds to
\begin{equation}\label{func12}
       \frac{ \sigma_{\rm W_i} }{1 {\rm ms} } = (5.5\pm 1.0) \left( \frac{\nu}{1 {\rm Ghz}} \right)^{-1.7 \pm 0.6}.
\end{equation}
Figure \ref{fig3}c shows the maximum and minimum values of the intrinsic pulse width in
each subsample. At a frequency of $\nu < 2$ GHz, the minimum width is about 0.1 ms, while
the maximum width can be up to $\sim 65$ ms. At 6 GHz, the minimum width is as short as
$\sim 3 \,\times\, 10^{-3}$ ms and the maximum width is $\sim 3$ ms. The maximum and minimum
widths can put stringent constraints on the trigger mechanism and radiation process of
FRBs. A successful FRB model should be capable of explaining these extreme values (\citealt{Beloborodov2020}; \citealt{Zhang2022in}).

\begin{figure}
        \centering
    \includegraphics[width=0.49\textwidth]{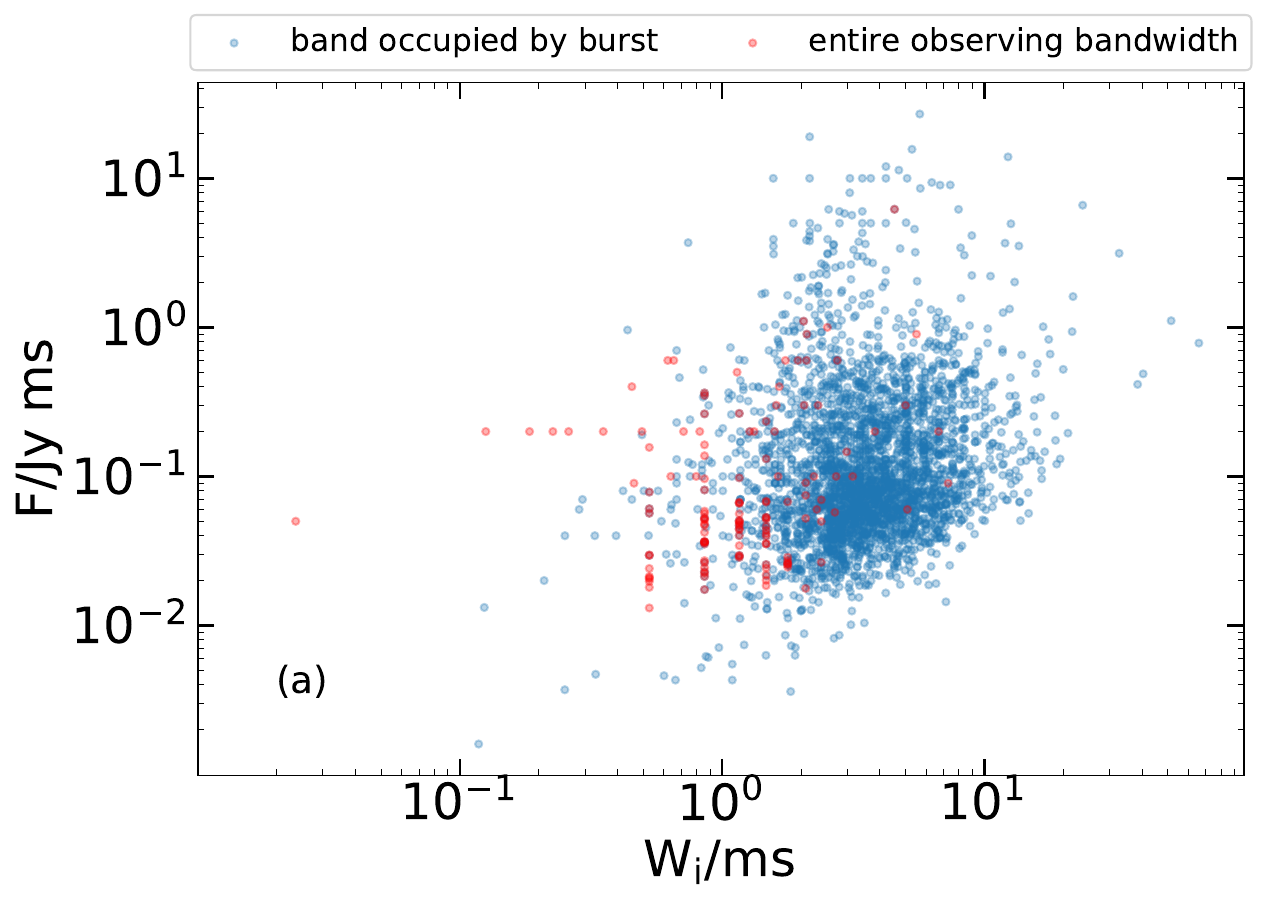}
        \includegraphics[width=0.49\textwidth]{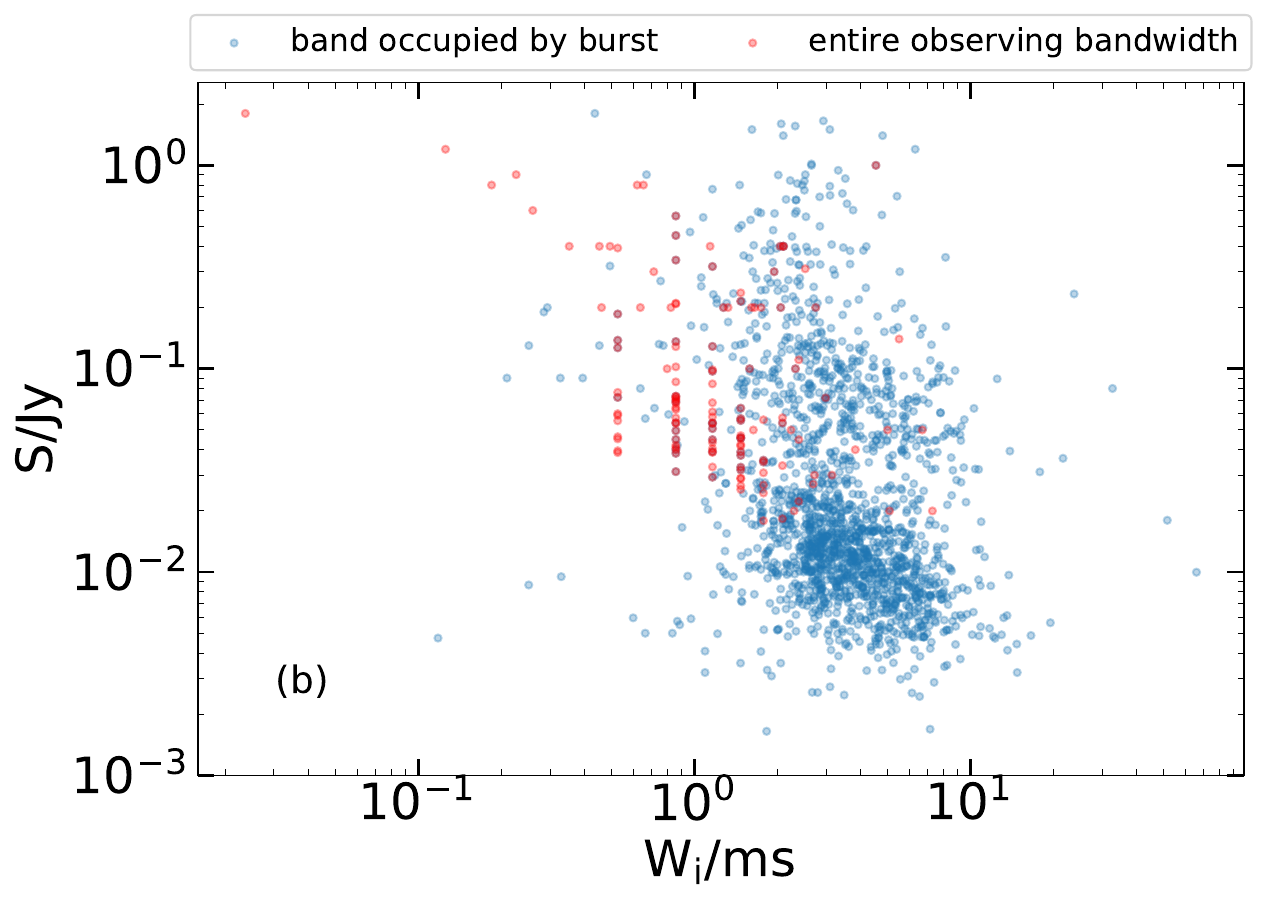}
        \includegraphics[width=0.49\textwidth]{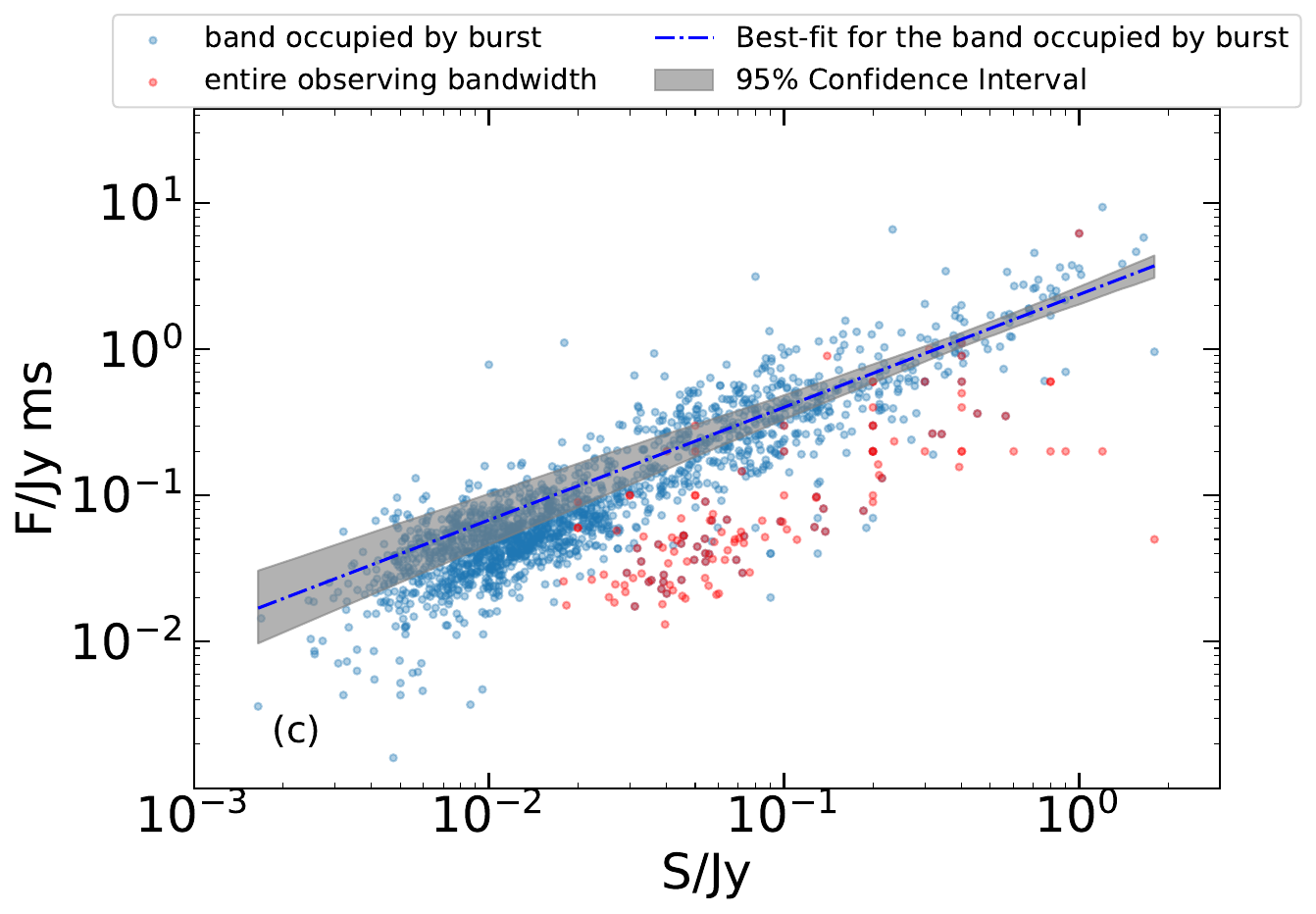}
        \caption{Overall features of the FRBs.
    Red points correspond to those bursts whose fluence is derived
    by integrating across the entire observing bandwidth. Blue points correspond
    to the events with fluence derived from the band occupied by
    the burst.
    Panel (a): FRBs on the $F$-$W_{\rm i}$ plane;
    Panel (b): FRBs on the $S$-$W_{\rm i}$ plane;
    Panel (c): FRBs on the $F$-$S$ plane.}
    \label{fig4}
\end{figure}

The peak flux ($S$), fluence ($F$) and pulse width are three key
parameters of FRBs. After acquiring the intrinsic pulse width, we
can explore any possible connection between the parameters. Figure
\ref{fig4} shows the distribution of FRBs when one parameter is
plotted versus another  for the bursts with fluence and
peak flux available.  Note that the methods used to
acquire the fluence are different by various groups so that there
could exist a systematic discrepancy between different subsamples.
Some authors calculated the fluence simply across the entire
observing bandwidth, while others derived the fluence from the
band occupied by the burst. To make a distinction, the data points
corresponding the former cases are plotted in red (which only
include 139 bursts) and the data in the latter cases are shown in
blue (over 3000 bursts).  From Figure \ref{fig4}a, we see that
the distribution of the FRBs on the $F$-$W_{\rm i}$ plane is quite
chaotic so that no correlation exists between these two
parameters. Another feature is that the pulse widths of the
majority of FRBs are distributed in a relatively narrow range
($\sim 1$ -- 10 ms). On the contrary, the distribution range of
fluence is much wider, which spans over two magnitudes ($\sim
0.01$ -- 1 Jy ms). Figure \ref{fig4}b shows that there seems to
exist a weak negative correlation between the peak flux and the
intrinsic pulse width: a longer FRB tends to have a smaller peak
flux. However, the correlation is generally too weak to draw any
firm conclusion. Similarly, we notice that the peak flux also has
a relatively large distribution range, which spans over two
magnitudes (mainly in a range of $\sim 3 \times 10^{-3}$ -- $3
\times 10^{-1}$ Jy). Figure \ref{fig4}c plots the fluence ($F$)
versus the peak flux ($S$). A clear positive correlation
can be seen in the plot. Since the red points correspond to a
minor portion of the data, we have analyzed the majoring blue
points separately. A best fit of the blue data points gives
\begin{equation}\label{func13}
    \frac{ F }{1 {\rm Jy \, ms} } = (2.4\pm 0.2) \left( \frac{S}{1 {\rm Jy}} \right)^{0.8\pm 0.1}.
\end{equation}
In fact, $F$ can be regarded as a product of the peak flux and the equivalent
pulse width. The above positive correlation is a natural outcome giving the
fact that the intrinsic pulse widths of FRBs are narrowly distributed in a
small range.

\begin{figure*}
        \centering
        \includegraphics[width=0.438\textwidth]{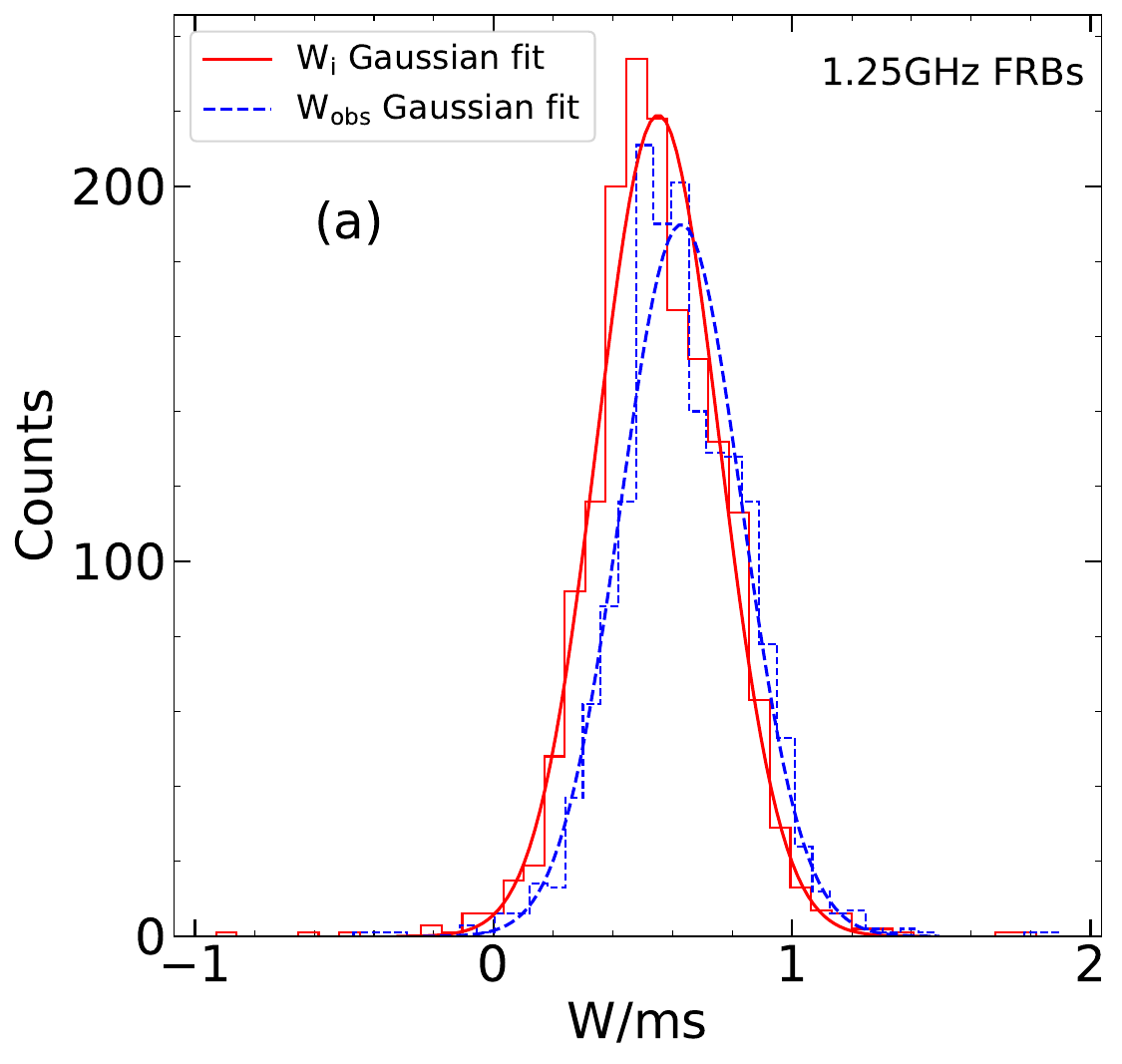}
        \includegraphics[width=0.438\textwidth]{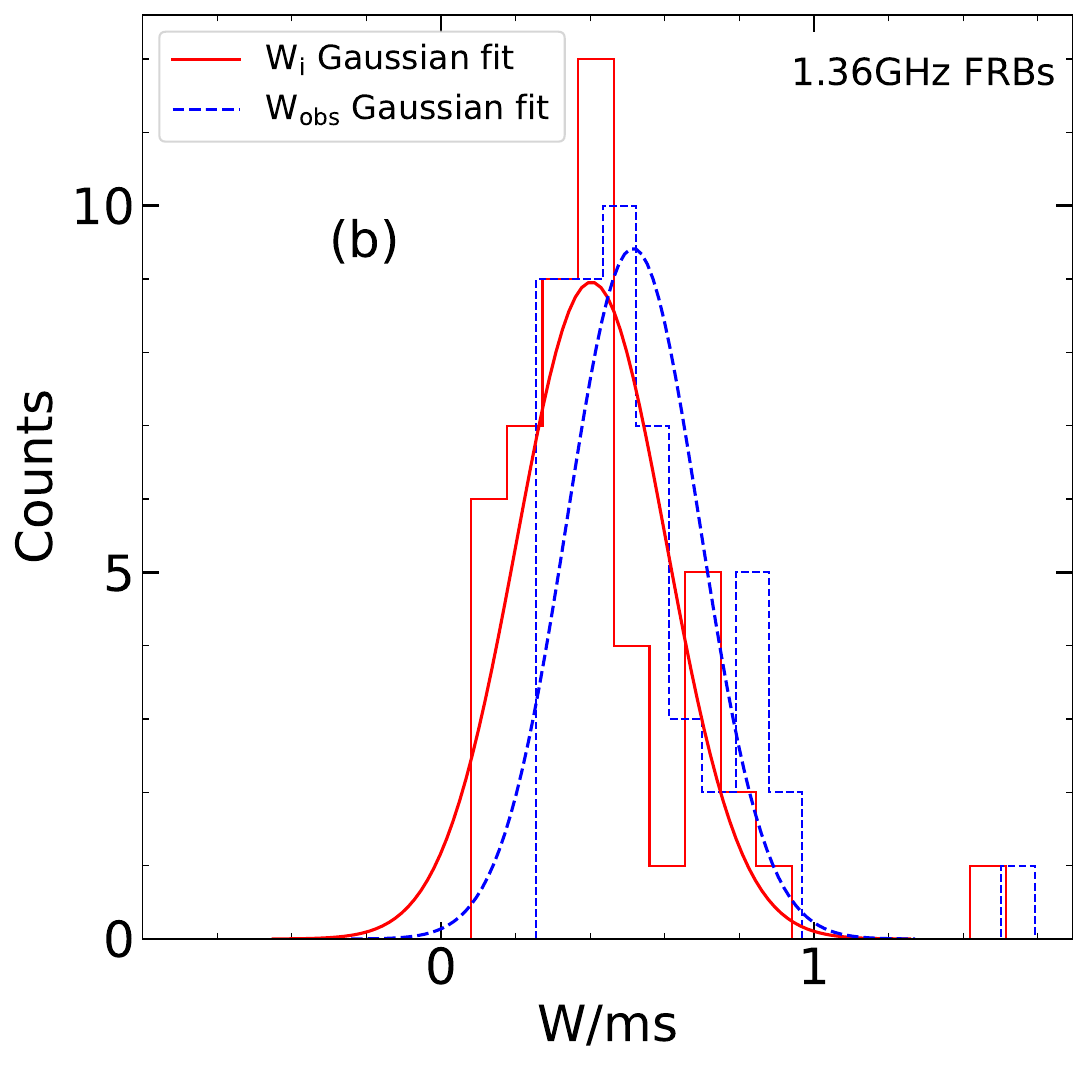}
        \includegraphics[width=0.438\textwidth]{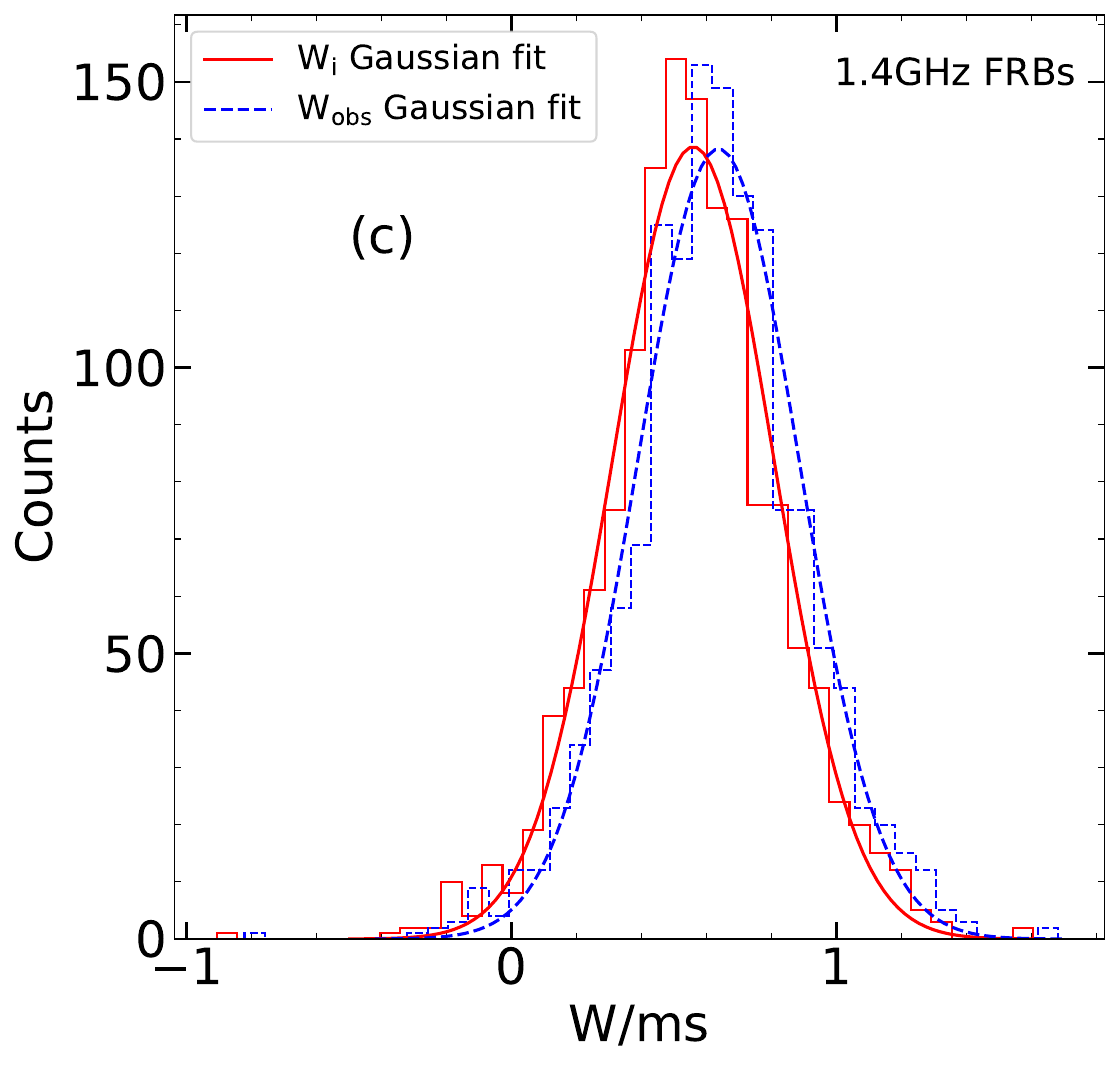}
      \includegraphics[width=0.438\textwidth]{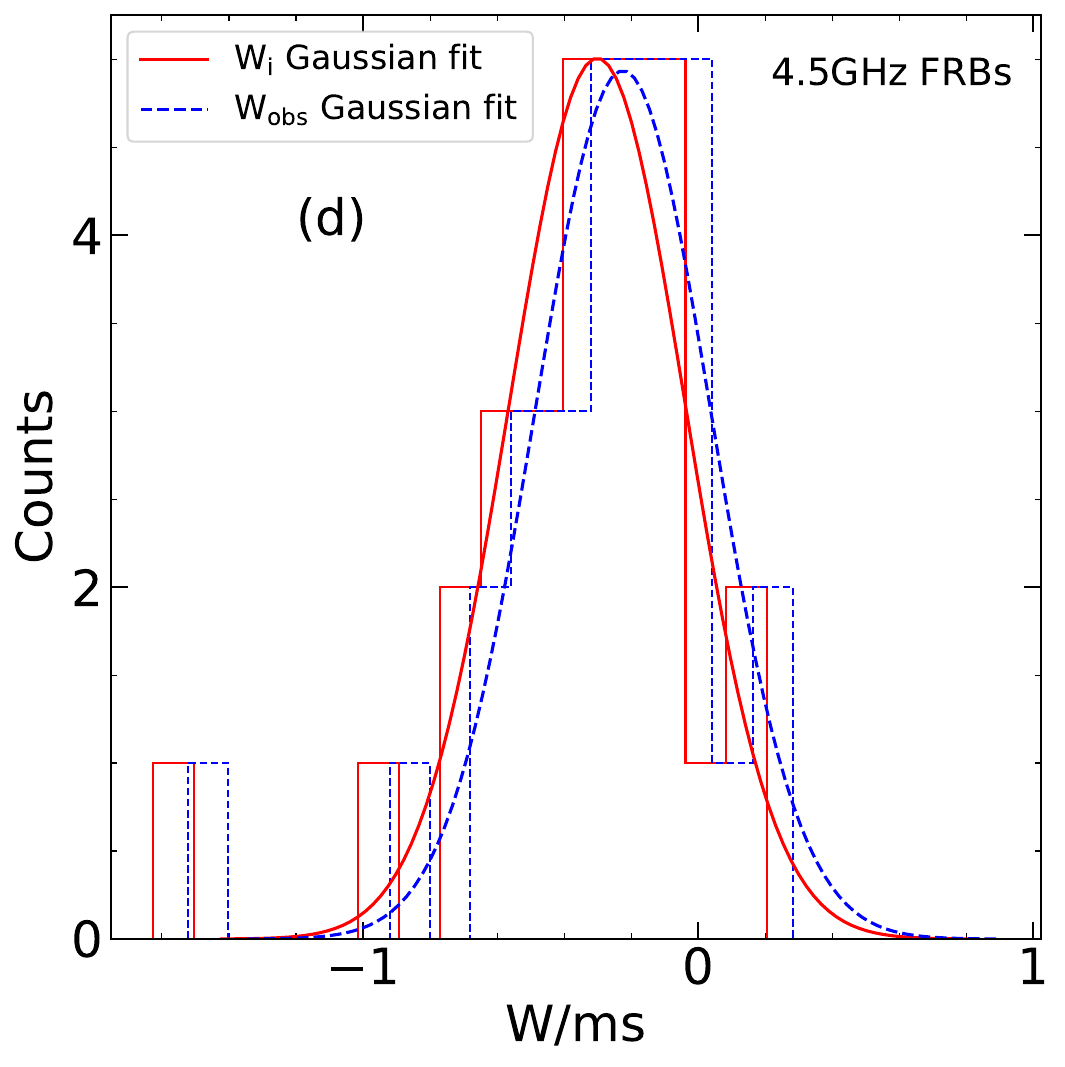}
        \includegraphics[width=0.438\textwidth]{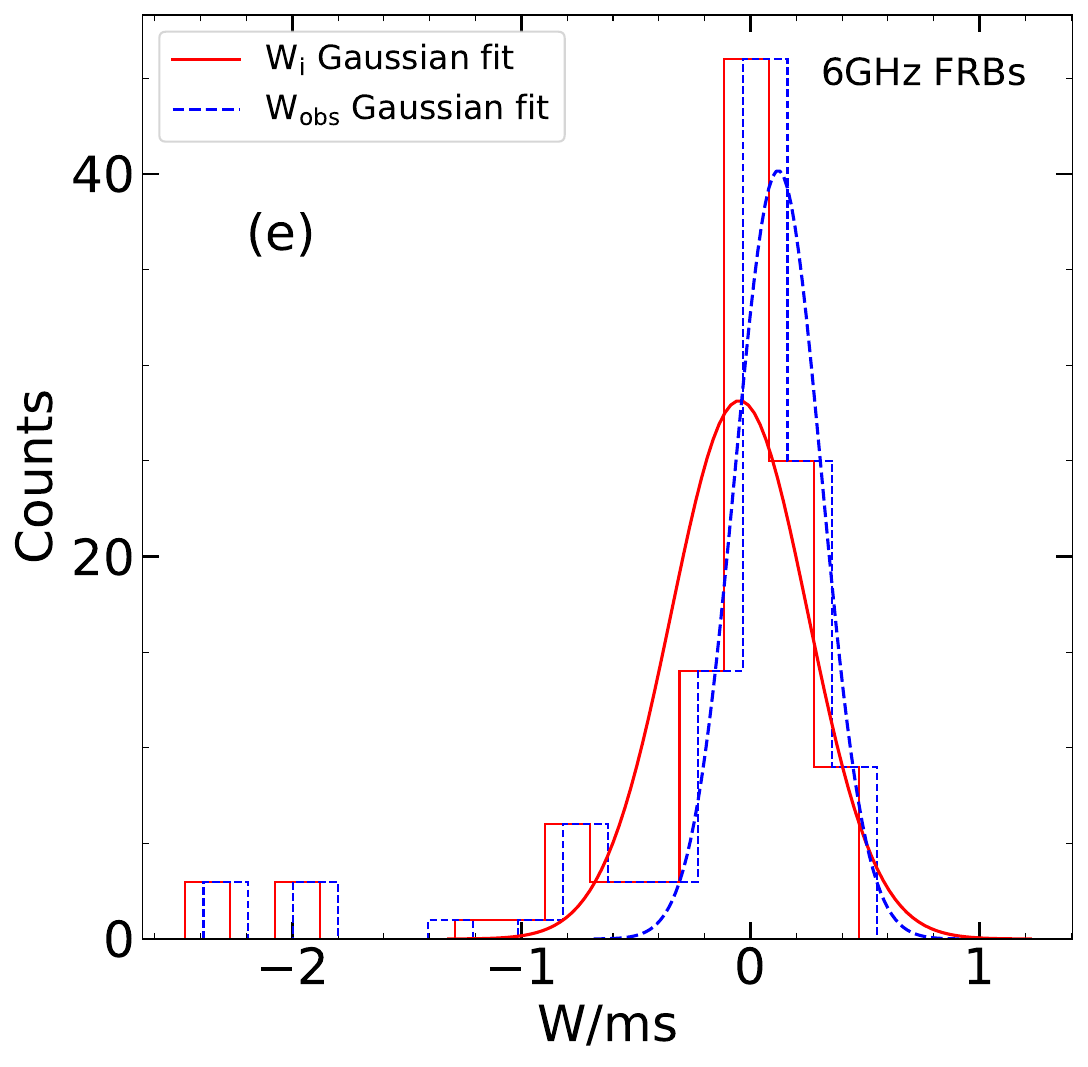}
        \includegraphics[width=0.438\textwidth]{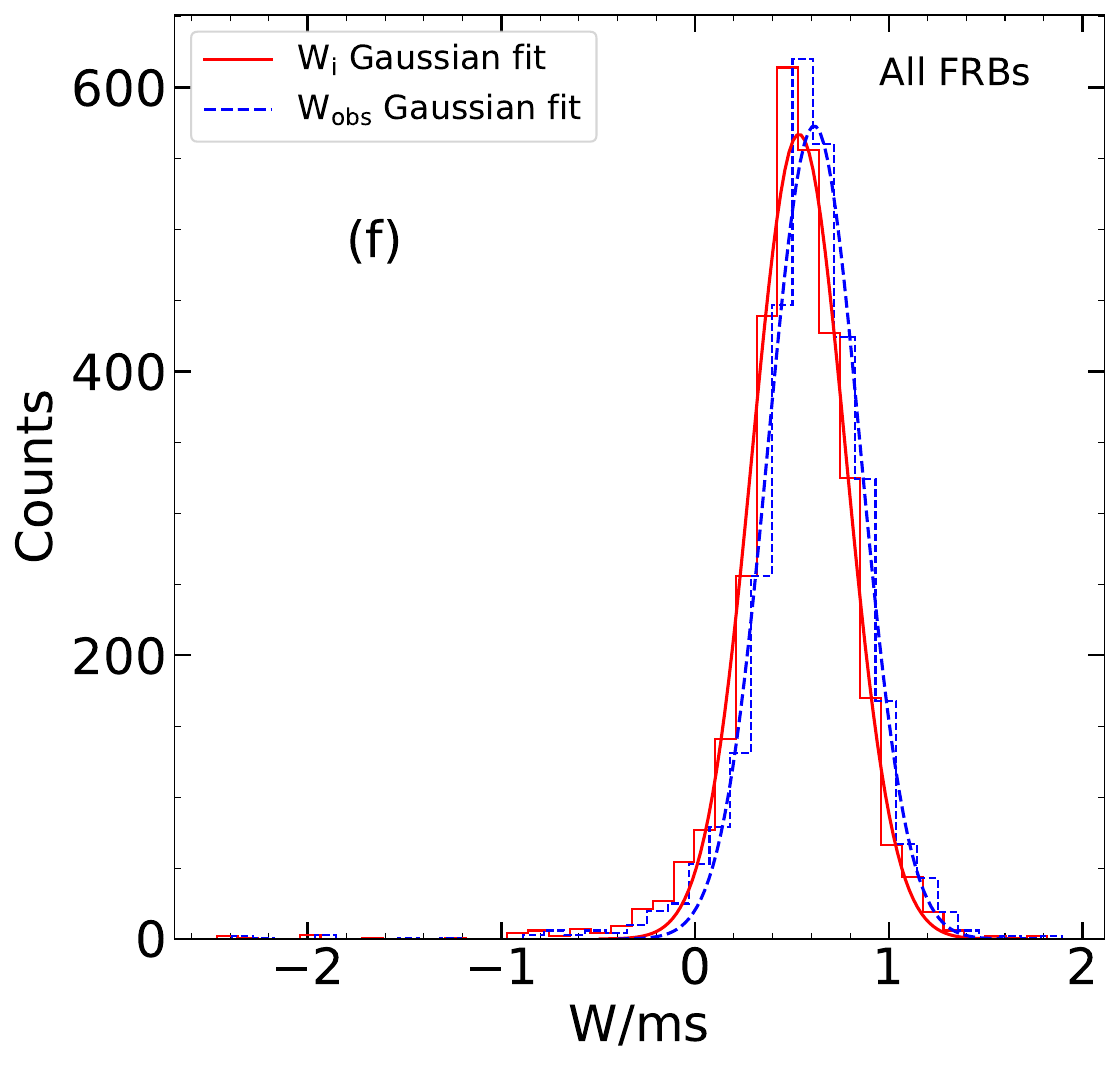}
        \caption{The distribution of the observed/intrinsic pulse width for the five groups
    centered at 1.25 GHz (Panel a), 1.36 GHz (Panel b), 1.4 GHz (Panel c),
    4.5 GHz (Panel d) and 6 GHz (Panel e).
    The corresponding distribution of all the bursts as a whole is also plotted
    in Panel (f). The dashed curves correspond to the observed pulse width distribution and
    the solid curves correspond to the intrinsic width distribution.
    A log-norm fit is performed on each histogram and is illustrated
    by the corresponding solid or dashed curve. }
\label{fig5}
\end{figure*}

We have also divided the FRBs into several groups according to their central
frequency (i.e. 1.25 GHz, 1.36 GHz, 1.4 GHz, 4.5 GHz and 6 GHz groups).
The distribution of the observed/intrinsic pulse width is compared
for each group to present a direct illustration of the broadening effects.
Figure \ref{fig5}  plots the histogram distributions of the observed pulse
width and the corrected intrinsic pulse width for each group.
A Gaussian fit is performed for each distribution, which gives the most probable
width ($\mu_{\rm i}, \mu_{\rm obs}$) and the dispersion
range ($\sigma_{\rm i}, \sigma_{\rm obs}$) of the bursts at these frequencies.
The derived parameters are listed in Table \ref{table2}.
The intrinsic pulse widths are slightly smaller than the observed width,
but their distributions do not differ from each other significantly.

\section{Conclusions and Discussion}
\label{sec:conclusions}

With more and more FRB sources being detected, it is necessary to
analyze the statistical features of their intrinsic parameters
systematically. In this study, we have collected 3287 bursts and
their parameters detected from FRB 20121102A as of October 2023.
Various broadening effects on the pulse width, including the
scattering broadening caused by turbulence and the instrumental
broadening induced during observations, are considered and
corrected. It is found that these broadening effects do not play a
dominant role in the majority of bursts from FRB 20121102A. The
instrumental broadening is mainly in a range of $10^{-2}$ -- 1 ms,
which contributes a fraction of $10^{-3}$ -- $10^{-1}$ to the
observed pulse width. The scattering broadening is even smaller,
which mainly ranges in $10^{-5}$ -- $10^{-3}$ ms and only
contributes a fraction of $10^{-5}$ -- $10^{-2}$ to the observed
pulse width. A quasi-linear correlation exists between the fluence
and the peak flux, which is mainly due to the narrow distribution
of the intrinsic pulse width for the majority of the bursts.

Interestingly, the mean intrinsic pulse width is found to decrease
as the central frequency increases, i.e. $\overline{W}_{\rm i}
\propto  \nu^{-1.2 \pm 0.2}$. Some authors have also attempted to
quantify the relationship between pulse width and frequency
(\citealt{Gajjar2018}; \citealt{Chamma2023}). Especially,
\citet{Chamma2023} pointed out that the observed width scales as
$W_{\rm obs} \propto \nu^{-1.77 \pm 0.07}$, which is steeper than
the power-law of the intrinsic width derived in this study. More
strikingly, the dispersion range of the intrinsic pulse width also
has a decreasing behavior, $\sigma_{\rm W_i} \propto \nu^{-1.7 \pm
0.6}$. We have normalized the dispersion of intrinsic
pulse width by dividing it with the mean intrinsic pulse width to
get the quantity of $\sigma_{\rm W_i}/\overline{W}_{\rm i}$, which
is a measure of the relative dispersion level. It is found that
$\sigma_{\rm W_i}/\overline{W}_{\rm i}$ still has a decreasing
trend with frequency in the range of 1.3 -- 6 GHz, i.e.
$\sigma_{\rm W_i}/\overline{W}_{\rm i} \propto \nu^{-0.4 \pm
0.3}$. In our sample, the duration of low-frequency bursts can be
either long or short. For example, the widths of FAST bursts are
in a wide range of 0.3 -- 78.5 ms. However, the high-frequency
bursts are all short ones. Therefore, at high-frequencies, the
distribution of the observed pulse widths is more concentrated,
leading to a lower value of $\sigma_{\rm W_i}/\overline{W}_{\rm
i}$ correspondingly. The lack of long bursts at high-frequencies
may reflect the intrinsic property of the central engine, but it
could also be caused by observational selection effects since long
bursts are relatively difficult to detect at high frequencies due
to the sensitivity constraint. 

The frequency-dependent behavior of $\overline{W}_{\rm i}$ and
$\sigma_{\rm W_i}$ may place useful constraints on the triggering
mechanism and radiation process of FRBs.  It could be a natural
outcome of the curvature radiation mechanism. For a particular
observer, when the curvature radius of the magnetic field line is
larger, the electron bunches will spend more time sweeping across
the line of sight, which means the observed pulse width will be
longer. Besides, the characteristic frequency of curvature
radiation is inversely proportional to the curvature radius. It
then naturally leads to the inverse correlation between pulse
width and central frequency. However, note that other radiation
mechanisms, such as the coherent Cherenkov radiation may also lead
to a similar behavior. According to \citet{LZN2023}, the
characteristic frequency of coherent Cherenkov radiation scales as
$\omega_{\rm ChR} \propto R^{-\frac{3}{2}}$, where $R$ is the
distance from the magnetar to the emission site. Again, a larger
$R$ will correspond to a longer pulse width since the curvature
radius is higher. More detailed analysis is necessary to further
discriminate between various mechanisms, which is beyond the scope
of this study.

It is worth noting that some bursts may have multiple
components, i.e. sub-bursts. However, until now, there has been no
consensus on how to distinguish between bursts and sub-bursts. For
example, \citealt{Jahns2023} distinguished bursts and sub-bursts
by eyes. \citealt{Li2021} argued that bursts with a complex
time-frequency structure should be regarded as sub-bursts when
there is a ``bridge'' emission between pulses.
\citealt{Hewitt2022} proposed a more stringent criterion. They
suggested that a pulse could be identified as a distinct burst
only when its Gaussian-like profile does not smear into other
components but instead begins and ends at a baseline comparable to
the noise. Otherwise, the components should be classified as
sub-bursts. Obviously, the criterion of identifying sub-pulses
will affect the study involving the pulse widths, which should be
paid more attention in future researches.


 Direct observational measurements of scintillation have
been attempted in some previous studies (\citealt{Spitler2016};
\citealt{Michilli2018}; \citealt{Gajjar2018};
\citealt{Hessels2019}). Using the uncertainty relation of $2 \pi
\tau_{\rm sc} \delta \nu_{\rm sc} = 1$ for a homogeneous medium or
$2 \pi \tau_{\rm sc} \delta \nu_{\rm sc} = 1.16$ for a Kolmogorov
medium, where $\delta \nu_{\rm sc}$ is the scintillation
bandwidth, the scattering timescale can also be derived from the
scintillation bandwidth correspondingly (\citealt{Cordes1998}).
For example, \citealt{Spitler2016} reported that the
scintillation bandwidth is 50 kHz at 1.5 GHz, which will lead to a
scattering timescale of $\sim 4 \times10^{-3}$ ms at this
frequency. \citealt{Hessels2019} reported the scintillation
bandwidth as 58.1 kHz at 1.65 GHz, leading to a corresponding
scattering timescale of $\sim 6\times10^{-3}$ ms.
\citealt{Michilli2018} argued that the scintillation bandwidth is
$\sim$ 5 MHz at 4.5 GHz, corresponding to a scattering timescale
of $\sim 3\times10^{-5}$ ms at that frequency.
\citealt{Gajjar2018} found that the scintillation bandwidth ranges
from 7 to 87 MHz at 4.5 - 8 GHz, which corresponds to a scattering
timescale of $\sim 6 \times10^{-6}$ ms in that frequency range.
The scattering timescales derived in our study ($10^{-5}$ --
$10^{-3}$ ms in 1.3 -- 6 GHz) are consistent with all those
previous results.

Although the scattering broadening seems to be insignificant in the case
of FRB 20121102A, it may still play an important role in other FRB sources.
In fact, from Equation (\ref{func2}), we see that the scattering broadening
scales as $\tau_{\rm sc}  \propto
\lambda_0^4 \mathrm{DM}^2 (\frac{\delta n_{\rm e}}{n_{\rm e}})^2 l_0^{-1}$. So, it
sensitively depends on many parameters, such as the observing frequency,
the total dispersion measure, the turbulence scale, and the density
fluctuations. Especially, when the central frequency decreases to 300 MHz,
the scattering broadening will be enlarged by a factor of $\sim 120$.
Note that the instrumental broadening may also be significantly amplified at low
frequencies (see Equation (\ref{func7})). As a result, these broadening effects
should still be paid attention to in other FRB sources as long as the pulse
width is involved.

Turbulence is also a complex factor that could seriously affect the two
ISM parameters in Equation (\ref{func2}), i.e. $l_0$ and $\delta n_{\rm e}$.
Additionally, note that the spectrum index of turbulence ($\beta$) may also
have some effects on the scattering broadening. In practice, when an FRB is
detected by a radio telescope, the dynamical spectrum (or
the so-called ``waterfall'' diagram) is usually available, which may contain rich
information on the turbulence of the ISM. A detailed analysis of the dynamical
spectrum may help determine these parameters.
On the other hand, we have mainly considered the situation where the
diffractive length of the ISM is smaller than $l_{0}$. In realistic cases,
the diffractive length could also be larger than $l_0$. It may lead to
some differences in the scattering broadening, which need to be addressed
in the future.

\section*{acknowledgements}

We would like to thank the anonymous referees for helpful
suggestions that led to an overall improvement of this study.
 Our work was supported by the National Natural Science
Foundation of China (Grant Nos. 12041306, 12233002, 12273113), by the
National SKA Program of China Nos. 2020SKA0120300 and
2022SKA0120102,
by the National Key R\&D Program of China (2021YFA0718500, 2023YFE0102300),
by the CAS ``Light of West China'' Program (No. 2021-XBQNXZ-005), by Xinjiang Tianshan Talent Program,
and by the CAS Project for Young Scientists in Basic Research (Grant No. YSBR-063).
YFH also acknowledges the support from the Xinjiang Tianchi Program.

\bibliography{sample631}{}
\bibliographystyle{aasjournal}

\begin{table*}
        \begin{center}
        \caption{An overview of the observations and the corresponding telescope parameters}
        \label{table1}
        \setlength{\tabcolsep}{3pt}
        \begin{tabular}{c c c c c c}
        \hline\hline\noalign{\smallskip}
        {Telescope}&{Central Frequency/GHz}&{$\Delta\nu_{\rm MHz}$}&{$\tau_{\rm samp}$/ms}&{Burst Counts}&{Ref.}\\
        \hline\noalign{\smallskip}
        FAST & 1.25& 0.122& 0.0983& 1652&(1)\\\hline
        \multirow{2}{*}{Arecibo}&{ 1.4}& {0.34/1.56}&{ 0.655/0.0102}& {11/1360}&
        (2),(3),(4),(5),(6),(7),(8)\\
                        &{ 4.5}& {1.56}& {0.0102}& {29}&(9),(10)   \\\hline
        \multirow{2}{* }{GBT}& {2}& {1.56}&{0.0102}&{19}&(5),(6),(11)     \\
                        &{ 6}&{0.183/0.366/2.930}& {0.0102/0.35/0.000341}& {2/93/19}&(9),(12),(13)\\ \hline
        \multirow{2}{* }{Effelsberg}&{ 1.36}& {0.586}& {0.0546}& {49}&(14),(15)\\
                        &{ 6}& {0.977}& {0.131}& {1}&(10)\\\hline
        VLA & { 3}& {0.25/4}& {1.024/5}& {2/9}&(10),(16)\\ \hline
        WSRT & {1.4}& {0.781/0.195}& {0.00128/0.0819}& {29/1}&(17)\\ \hline
        MeerKAT &{ 1.28}& {0.209}&{ 0.00479}& {11}&(18)\\ \hline
\end{tabular}
\end{center}
\begin{tablenotes}
     \footnotesize
      \item \textbf{Ref.} (1) \citealt{Li2021}; (2) \citealt{Spitler2016}; (3) \citealt{Hewitt2022}; (4) \citealt{Jahns2023}; (5) \citealt{Hessels2019}; (6) \citealt{Scholz2017}; (7) \citealt{Gourdji2019}; (8) \citealt{MAGIC2018}; (9) \citealt{Michilli2018}; (10) \citealt{Hilmarsson2021}; (11) \citealt{Scholz2016}; (12) \citealt{ZhangY2018}; (13) \citealt{Snelders2023}; (14) \citealt{Cruces2021}; (15) \citealt{Hardy2017}; (16) \citealt{Law2017}; (17) \citealt{Oostrum2020}; (18) \citealt{Caleb2020}.
     \end{tablenotes}
\end{table*}

\begin{table}
\begin{center}
\caption{The best-fit parameters for the log-normal distribution of bursts in each frequency group}
\label{table2}
\setlength{\tabcolsep}{7pt}
\begin{tabular}{c c c c c }
\hline
\hline
Central Frequency/GHz&\textbf{$\mu_{\rm i}$/ms}&\textbf{$\sigma_{\rm i}$/ms}&\textbf{$\mu_{\rm obs}$/ms}&\textbf{$\sigma_{\rm obs}$/ms}\\
\hline
All             & 3.47& 1.74& 4.07 & 1.74\\
1.25         & 3.55 & 1.58 & 4.27 & 1.58\\
1.36         & 2.51 & 1.58 & 3.24 & 1.51 \\
1.4         & 3.63 & 1.78 & 4.37 & 1.78\\
4.5         & 0.50 & 1.82 & 0.60 & 1.82\\
6            & 0.89 & 2.00 & 1.32 & 1.55\\
\hline
\end{tabular}
\end{center}
\end{table}

\end{CJK*}
\end{document}